\documentclass[aps,prb,amsmath,amssymb,amsfont,showpacs,superscriptaddress,reprint]{revtex4-1}
\usepackage{amsmath, amssymb, amsfonts, multirow}
\usepackage{bm}
\usepackage[dvipsnames]{xcolor}
\usepackage{braket}
\usepackage[artemisia]{textgreek}  
\usepackage{float}
\usepackage[version=3]{mhchem}
\usepackage{tabularx}
\usepackage{graphicx}

\begin{document}

\title{Improved alkali intercalation of carbonaceous materials in ammonia solution}

\author{B. G. M\'{a}rkus}
\affiliation{Department of Physics, Budapest University of Technology and Economics and MTA-BME Lend\"{u}let Spintronics Research Group (PROSPIN), P.O. Box 91, H-1521 Budapest, Hungary}

\author{S. Kollarics}
\affiliation{Department of Physics, Budapest University of Technology and Economics and MTA-BME Lend\"{u}let Spintronics Research Group (PROSPIN), P.O. Box 91, H-1521 Budapest, Hungary}

\author{P. Szirmai}
\affiliation{Institute of Physics of Complex Matter, FBS Swiss Federal Institute of Technology (EPFL), CH-1015 Lausanne, Switzerland}

\author{B. N\'afr\'adi}
\affiliation{Institute of Physics of Complex Matter, FBS Swiss Federal Institute of Technology (EPFL), CH-1015 Lausanne, Switzerland}

\author{L. Forr\'o}
\affiliation{Institute of Physics of Complex Matter, FBS Swiss Federal Institute of Technology (EPFL), CH-1015 Lausanne, Switzerland}

\author{J. C. Chac\'on-Torres}
\affiliation{Yachay Tech.~University, School of Physical Sciences and Nanotechnology, 100119 Urcuqu\'{i}, Ecuador}

\author{T. Pichler}
\affiliation{Faculty of Physics, University of Vienna, Strudlhofgasse 4, A-1090 Vienna, Austria}

\author{F. Simon}
\affiliation{Department of Physics, Budapest University of Technology and Economics and MTA-BME Lend\"{u}let Spintronics Research Group (PROSPIN), P.O. Box 91, H-1521 Budapest, Hungary}

\keywords{intercalation, doping, liquid ammonia solution, charge transfer, Raman spectroscopy, ESR}

\begin{abstract}
Alkali intercalated graphite compounds represent a compelling modification of carbon with significant application potential and various fundamentally important phases. We report on the intercalation of graphite with alkali atoms (Li and K) using liquid ammonia solution as mediating agent. Alkali atoms dissolve well in liquid ammonia which simplifies and speeds up the intercalation process, and it also avoids the high temperature formation of alkali carbides. Optical microscopy, Raman, and electron spin resonance spectroscopy attest that the prepared samples are highly and homogeneously intercalated to a level approaching Stage-I intercalation compounds. The method and the synthesis route may serve as a starting point for the various forms of alkali atom intercalated carbon compounds (including carbon nanotubes and graphene), which could be exploited in energy storage and further chemical modifications.
\end{abstract}

\maketitle

\section{Introduction}
Modifying the charge state in the various forms of carbon has lead to important applications and to the discovery of appealing correlated phases. The examples include the now widespread application of lithium-intercalated graphite electrode for energy storage \cite{LiGraphite_Review}, the $T_{\text{c}}=11.5$ K superconductor CaC$_6$ (Ref. \cite{EmeryCaC6}), alkali intercalated fullerides with a relatively high $T_{\text{c}}=29$ K (for Rb$_3$C$_{60}$ \cite{RosseinskyRb3C60}). The modified electronic structure due to the charge transfer helped to understand the optical properties of single-wall carbon nanotubes \cite{Eklund1997} and helped to unravel the correlated ground state by the discovery of the Tomonaga--Luttinger to Fermi liquid state \cite{KatauraNAT2003,PichlerPRL2004}. 

The most conventional synthesis method to obtain alkali intercalated carbon compounds is the so-called two zone vapor phase method, which works well for heavier alkali atoms with a lower melting point (K, Rb, and Cs) \cite{Dresselhaus1981}: the alkali atoms are heated together with the desired form of carbon (graphite, graphene, nanotubes, or fullerene). For Li and Ca doping, the intercalation of graphite was achieved by immersing the sample into molten Li or Li/Ca mixtures \cite{EmeryCaC6} with temperatures up to $350~^{\circ}$C. This, relatively high temperature is dictated by the melting point and due to the kinetics of the diffusive process. However, this method leaves a relatively small reaction window for the intercalation, as formation of alkali carbides starts at higher temperatures (around $450~^{\circ}$C) and also requires bulk, crystalline samples. An alternative route to alkali atom doping proceeds with the help of solvents, such as liquid ammonia and organic solvents (e.g. methylammonium, CH$_3$NH$_2$ or tetrahydrofuran (THF)). These are known to dissolve the alkali and some alkaline earth elements well. Then, the reaction between the alkali atoms and the carbon material proceeds in the solution at moderate temperatures and with a high efficiency, due to the large reaction surface. This procedure was used to yield highly alkali doped fullerides \cite{MurphyJPCS1992,BuffingerJACS1993,DahlkeJACS2000,AjayanACSNano2011,PrassidesCs3C60Nature,hirsch} and carbon nanotubes \cite{SzirmaiPRB}. It is expected that by conducting a liquid ammonia intercalation procedure, one could also facilitate the synthesis of alkali intercalated graphite, which could be the starting material for functionalized graphite, graphene or nanoribbons.

Here, we report a one-step synthesis of highly alkali intercalated graphite using liquid ammonia. This process does not involve the exposure to ambient condition once the intercalation started and the samples are directly measurable afterwards. The whole process is conducted under a closed vacuum system, thus the amount of toxic ammonia can be minimized, moreover oxidation of alkali is suppressed. We use microscopy, Raman and electron spin resonance (ESR) spectroscopic techniques to prove that the resulting materials are indeed highly intercalated. We also thoroughly investigate well known side reaction species (such as amides and imides) and prove using ESR that these are absent in our final material.

\section{Intercalation in ammonia solution}

The doping or intercalation process is based on the fact that liquid ammonia, similarly to water is in a self-ionized (autodissociated) state \cite{chambers1975,greenwood1998}:
\begin{equation}
\ce{2 NH_3 <=> NH_4+ + NH_2-}.
\end{equation}
Introducing alkali or alkaline earth metals (denoted with $\ce{M_{am}}$) into the liquid the metal gets ionized and the electron will be donated to the ammonia molecule:
\begin{gather}
\ce{M_{am} -> M_{am}+ + e-}, \\
\ce{e- + $x$NH_3 -> e$^{-}$(NH_3)_{$x$}}.
\end{gather}

The alkali ammonia solution has a dark blue color, when the metal concentration is low, while a golden yellow or brownish color is realized, when the concentration of the metal is high \cite{greenwood1998}. The alkali ions in the liquid can freely move with diffusion and can penetrate the host material. At the end of the intercalation process the electron is donated back from the ammonia, which results in a doped material. Since the electronegativity of the alkali and alkaline earth metals is usually weak the electron is thus donated to the host material.

\subsection{Preparation}

Before the intercalation or doping could take place, several preparatory steps are mandatory. The host material and the intercalant metal is placed into the quartz tube in an argon-filled glovebox (MBraun GmbH with $< 0.1$ ppm of O$_2$ and H$_2$O) and sealed with a vacuum valve. To maximize the reaction efficiency special sphere ended quartz tubes were used (depicted in the inset of Fig. \ref{fig:nh3intercal}). The diameter of the ending bubble was $10-15$ mm, while the diameter of the quartz tube was $5$ mm. In principle the bigger volume of the sphere allows bigger amounts to be simultaneously doped as the reaction surface is increased. In the current work Grade-I highly oriented pyrolytic graphite (HOPG) discs from SPI Supplies with a diameter of $3$ mm and a thickness of $70~\mu$m, metallic lithium (Aldrich 01328TE $99.9+$$\%$) and potassium (Aldrich MKBL0124V $99.9+$$\%$) from Sigma-Aldrich were used. Here we point out that the method is not limited to graphite host, as it was proven earlier for C$_{60}$ and should work with other materials as well (e.g. SWCNTs, graphene, black phosphorus, etc.) and should work with any ammonia soluble metal (e.g. alkali and alkaline earth metals). The amount of intercalant was in a minor excess regarding the stoichiometry required for a Stage-I compound (LiC$_6$ and KC$_8$). Samples prepared with the current setup are typically in the scale of a few milligrams.

\subsection{Reaction}

\begin{figure}[h!]
\includegraphics*[width=\linewidth]{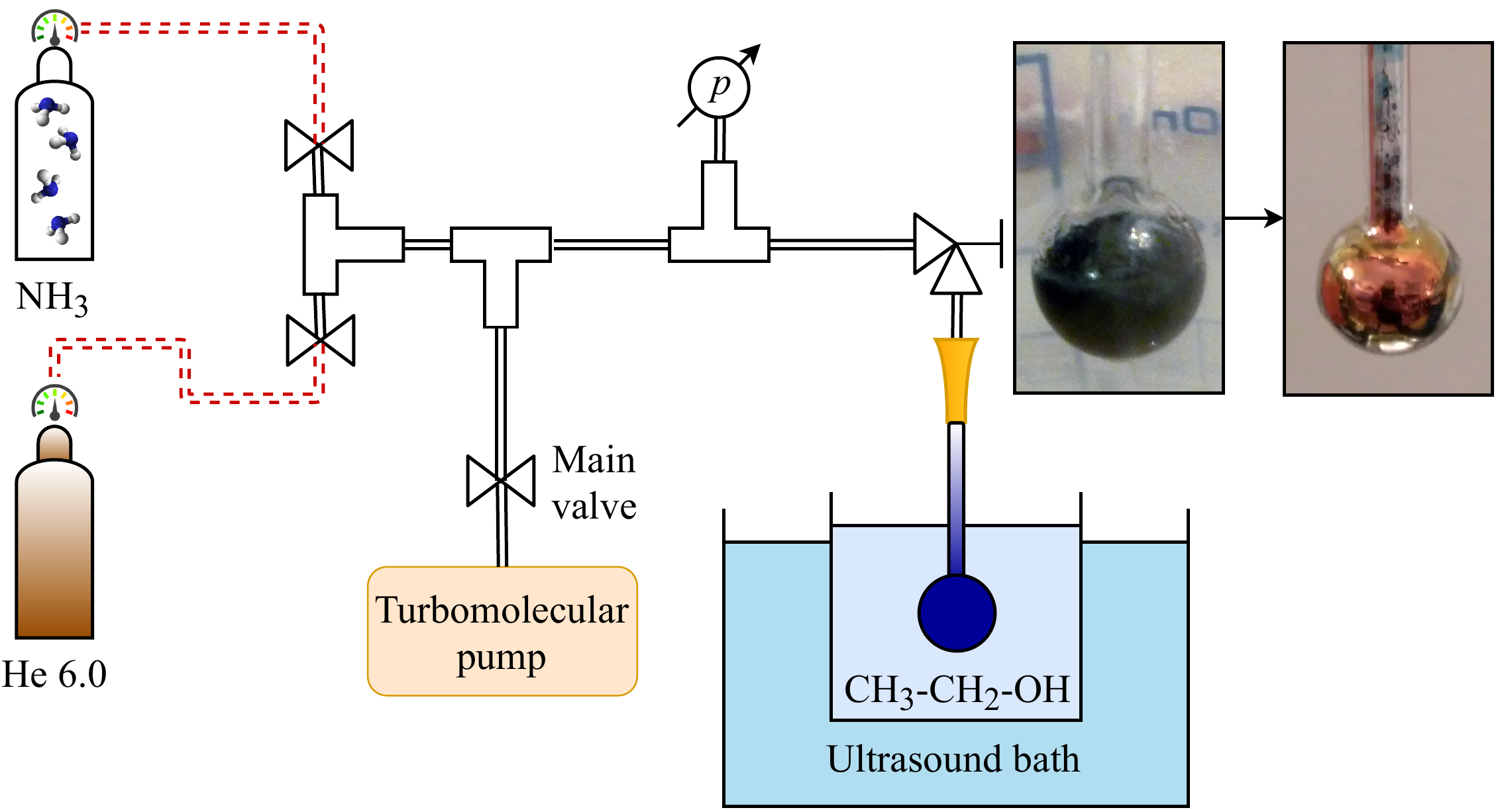}
\caption{Schematic diagram of the used intercalation setup, connected to a vacuum setup with gas inlets. The solution is represented with a blue bubble connected to the system. Underneath a plastic bottle filled with cooled ethanol is present, which keeps the temperature of the solution below the boiling point of the ammonia. The ethanol is cooled with periodically introduced liquid nitrogen. The intercalation is assisted with a bath sonicator placed under. The pressure of the ammonia and helium gases are measured with a broad range pressure meter. Valves are represented with triangles facing each other. Inset shows the solution at the beginning and at the end of the process. The solution has a dark blue color, when the concentration of the metal is low, and yellowish-brownish when high.}
\label{fig:nh3intercal}
\end{figure}

After the host and the intercalant are placed in the quartz tube, it is connected to a vacuum system shown schematically in Fig. \ref{fig:nh3intercal}. Here the argon gas is removed and the system is evacuated to high vacuum ($2 \times 10^{-6}$ mbar).

Closing the main valve to the turbomolecular pump and introducing $900-950$ mbar of ammonia the reaction can begin. Here the ammonia has to be condensed rapidly with liquid nitrogen, because the ammonia gas also reacts with the alkalis, as shown later. The drop of the ammonia gas pressure to about $400-500$ mbar (depending on the size of the total system) at this point is normal, and it is due to ammonia liquefaction. Underneath the now frozen sample a readily cooled ethanol ($-60~^{\circ}$C) and a bath sonicator is placed. The solid ammonia thus slowly melts and remains in the liquid solution phase avoiding the additional side reactions. At this point the alkali-ammonia system has a dark blue color, as shown on the right side of Fig. \ref{fig:nh3intercal}. The bath sonicator helps the diffusion of alkali ions into the host. Due to reaction kinetics, the solution temperature has to be kept close to the boiling point of the ammonia, typically between $-50~^{\circ}$C and $-40~^{\circ}$C. Here extra care has to be taken to avoid boiling of the ammonia, which happens at around $-33~^{\circ}$C. The whole intercalation process with sonication requires about $30$ minutes. At the end of the process the solution is slowly heated up, and its color turns into yellowish-brownish as presented in the right inset of Fig. \ref{fig:nh3intercal}. Once the ammonia gas pressure is restored to about $800-900$ mbar the gas is removed through the vacuum pump. Afterwards the sample is further sonicated for $30$ minutes in $30~^{\circ}$C water to remove the adsorbed ammonia and to avoid the sample being stuck to the inner wall of the quartz bubble. At the end the quartz is heated up to $200~^{\circ}$C for $30-60$ minutes to remove the rest of the ammonia and the excess alkali. Additionally, after closing the main valve a $20-30$ mbar of high purity He exchange gas can be introduced to make cryogenic measurements readily available without the need of placing the sample into a fresh tube. Finally the quartz tube is permanently sealed with a torch. Here we wish to emphasize that the sample is never in touch with oxygen or moisture, resulting a material free of oxides and oxidated species. Once the quartz ampule is sealed it is readily measurable, even down to a couple Kelvins with helium exchange gas present.

\subsection{Side reaction of gaseous ammonia and alkali metals} The clear drawback of the intercalation carried out in ammonia is that gaseous ammonia also reacts with the alkali metals forming alkali amides \cite{Titherley1894}. It worsens the situation that the reaction is catalytically activated, when iron, nickel, cobalt, platinum or their compounds are present \cite{Vaughn1934}. These are usual contaminants in graphite and graphene \cite{Sepioni2012}, and Fe or Ni catalyst particles are used in carbon nanotube growth. During the reaction, alkali amide salts are formed, which have white color and are usually insoluble in liquid ammonia:
\begin{equation}
\ce{2 M_{am,(s)} + 2 NH_{3(g)} -> 2 $\underset{\text{white}}{\ce{M_{am}NH_{2(s)}}}$ + H_{2(g)}$\uparrow$}.
\end{equation}
Here it is worth to mention that the resulting amide is not necessarily white, it can have various colors (pink or greenish-brownish) depending on other contaminants in the metal \cite{GayLussac1811}. For the case of lithium, if the amide concentration is high, lithium imide is formed, which has red color, and ammonia gas is released \cite{Michigoe2011}:
\begin{equation}
\ce{2 LiNH_{2(s)} ->[\ce{\text{decomposition}}] $\underset{\text{red}}{\ce{Li_2NH_{(s)}}}$ + NH_{3(g)}$\uparrow$}. 
\end{equation}

At higher temperatures (typically above $300~^{\circ}$C), other reaction routes are also opened, e.g., nitride and hydride formation. To avoid these, it is recommended not to go above $200~^{\circ}$C in the final step of the process.

Fortunately, most of these side reactions can be suspended or can be slowed down compared to the reaction time in the liquid phase. In the case of lithium, the amide formation would require $2-3$ weeks, for sodium $2-3$ days in the solution. Potassium is a bit more reactive and forms its amide within $2-3$ hours in the solution, even without the presence of the catalyst \cite{Franklin1905,Bergstrom1933}. Thus, the rapid cooling of the ammonia gas is essential.

\section{Intercalated HOPG prepared by the alkali-ammonia method}

We demonstrate the above mentioned intercalation process on highly oriented pyrolytic graphite (HOPG) discs. After the doping the color of the disks changed dramatically, proving that the intercalation was indeed successful and the final product is a graphite intercalated compound (GIC). We find that lithium intercalated samples have a golden-yellow color, thus, identified as Stage-I LiC$_6$. The potassium doped samples are mostly red with some yellowish parts, which can be interpreted as a mixture of Stage-I and Stage-II systems. It is known that there exists a metastable state of alkali intercalated graphite with the approximate stoichiometry of AC$_{10}$ \cite{Salzano1967}, which mostly has a red color. Our observations regarding the color change are experimentally verified with Raman and ESR spectroscopy. 

\paragraph{Sample characterization}

Raman measurements were carried out in a high sensitivity single monochromator LabRam spectrometer \cite{Fabian2011} using $514$ nm laser excitation, $50\times$ objective with $0.5$ mW laser power to avoid deintercalation. ESR measurements are performed on a Bruker Elexsys E500 X-band spectrometer at room temperature. Photographs were taken with Nikon Eclipse LV150N optical microscope both in bright and dark field with a $100\times$ objective.

\subsection{Raman spectroscopy}

\begin{figure}[h!]
\includegraphics*[width=.95\linewidth]{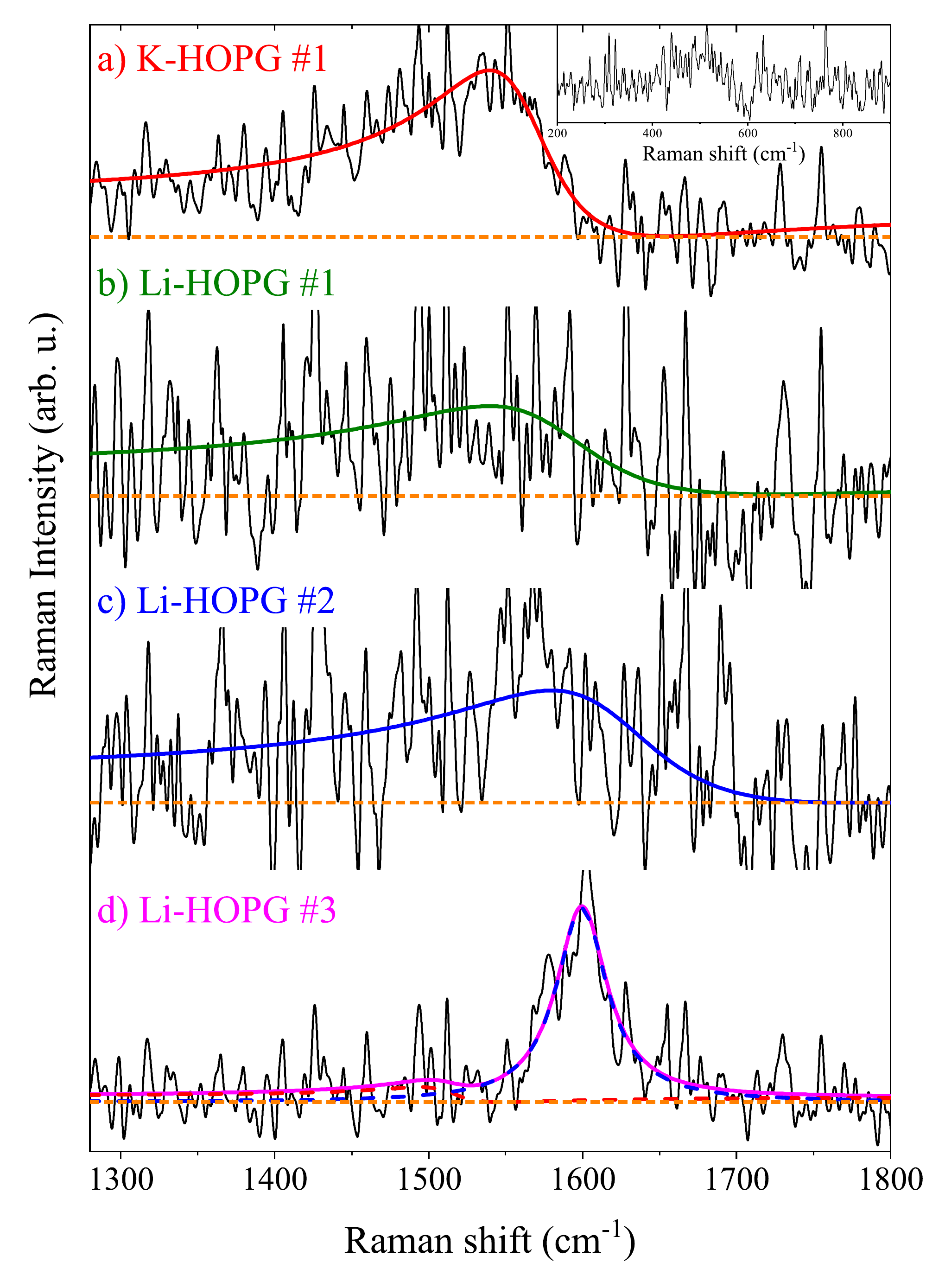}
\caption{Raman spectra of Li and K intercalated HOPG in the vicinity of the D and G modes of graphite. For samples a, b, and c only one Fano line was resolved denoting that the electrons and phonons are strongly coupled \cite{Fano1961}, and suggesting a high degree of intercalation. In the last spectrum, the doping of that sample is not complete, which is resulted in a mixture of a Fano line and a single Lorentzian, upshifted compared to the G-mode of the graphite. Similar behavior is observed for slightly dedoped LiC$_6$ material \cite{ChaconPRB2012,ChaconPSSB2012}. The mean value of the background signal is indicated with dashed orange line. Inset shows the characteristic C$_{\text{z}}$ mode of the potassium intercalated system. Here it is noted, that the strong noise of the spectra is due to thick quartz tube. {\color{black}During the measurements the samples were in the same quartz tube in which the reaction took place, to avoid degradation of the produced materials.}}
\label{fig:raman}
\end{figure}

Raman spectra near the D and G modes of host graphite of the prepared samples are presented in Fig. \ref{fig:raman}. Despite all our efforts, the resolution is limited by the thick quartz sample holder and the surface roughness of the samples due to the boiling ammonia at the end of the reaction. {\color{black}Here we wish to mention, that the samples were measured in the very same quartz tube in which the reaction took place. Replacing air sensitive materials is always a risky procedure, even in a glove box and we wanted to show that this step is not necessary with the current method, thus degradation of the produced materials can be avoided.} Nevertheless, the expected Fano lineshapes \cite{Fano1961} are observed for the well-intercalated species. On the other hand, we were unable to resolve the two $E_{2g}$ modes separately, similarly to previous results \cite{Eklund1977,Nemanich1977,Solin1980,Dresselhaus1977}.

In the Raman spectra of the investigated materials we observed a dominant Fano line in the vicinity of the G-mode of graphite. The presence of this shape is a clear indication that the electrons and phonons are strongly coupled, which is a result of high electron density \cite{Fano1961}. We address this to a successful intercalation of the host material to Stage-I. The parameters extracted from the curve fitting for the highly intercalated samples (spectrum a), b), and c) of Fig. \ref{fig:raman}) are summarized in Table \ref{tab:raman}.

\begin{table}[h!]
    \begin{center}
        \begin{tabular}{ccccc}
           \hline
           \hline
           Sample   & $\omega_{\text{Fano}}$ (cm$^{-1}$) & $\Gamma$ (cm$^{-1}$) & $q$     & $\gamma^{\text{EPC}}$  \\ \hline
           K $\#1$  & $1561$                             & $45.6$               & $-2.05$ & $191$ \\
           Li $\#1$ & $1581$                             & $77.1$               & $-1.84$ & $30$\\
           Li $\#2$ & $1618$                             & $78.1$               & $-2.02$ & $197$\\
           \hline
           \hline
        \end{tabular}
    \end{center}
    \label{tab:raman}
\end{table}

The parameters of the observed lines agree well with the previous literature observations: for KC$_8$ these are $1547$ \cite{Eklund1977,Nemanich1977}, $\sim 1500$ \cite{Solin1980}, $1522$ \cite{Saitta2008}, $1510$ and $1547$ \cite{ChaconPRB2012,ChaconPSSB2012,ChaconPSSB2011} for the position in cm$^{-1}$. The width of the peak is $78.7$ \cite{Eklund1977,Nemanich1977}, $157$ \cite{Saitta2008}, $176.7$ and $82.6$ \cite{ChaconPSSB2011}, $125$ and $70$ \cite{ChaconPSSB2012}, $125.6$ and $70.9$ \cite{ChaconPRB2012} in cm$^{-1}$. Notably, the position of the C$_{\text{z}}$ mode around $560$ cm$^{-1}$ also agrees well \cite{Eklund1977,Nemanich1977,Solin1980,Dresselhaus1977,ChaconPRB2012,ChaconPSSB2012,ChaconPSSB2011,Solin1981}. For the case of LiC$_6$, the literature values for the position are: $1594\pm3$ \cite{Eklund1980}, $1595$ \cite{Saitta2008}, $\sim1585$ \cite{Inaba1995}, $1546$ and $1585$ \cite{ChaconPRB2012,ChaconPSSB2012}, $1575$ and $1600$ \cite{Sole2014} in cm$^{-1}$. Here, the width is $48\pm4$ \cite{Eklund1980}, $70$ \cite{Saitta2008}, $71.0$ and $70.9$ \cite{ChaconPRB2012}, $71$ and $70$ \cite{ChaconPSSB2012}, $10$ and $\sim30$ \cite{Sole2014} in cm$^{-1}$.

The electron-phonon coupling parameter (EPC), which is the measure of the charge transfer can be calculated from the obtained parameters. The electron-phonon scattering linewidth can be estimated from the positions of the Fano lineshape using the expression:
\begin{equation}
\gamma^{\textrm{EPC}} \simeq 2\sqrt{(\omega_{\textrm{Fano}} - \omega_{\textrm{A}})(\omega_{\textrm{NA}}-\omega_{\textrm{Fano}})}.
\end{equation}
Here, $\omega_{\textrm{Fano}}$ is the measured position, $\omega_{\textrm{A}}$ and $\omega_{\textrm{NA}}$ are the calculated adiabatic and non-adiabatic phonon frequencies \cite{ChaconPRB2012,Saitta2008}. Here, the relevant quantities are for KC$_8$: $\omega_{\textrm{A}}=1223$ cm$^{-1}$, $\omega_{\textrm{NA}}=1534$ cm$^{-1}$, and $\omega_{\textrm{A}}=1362$ cm$^{-1}$, $\omega_{\textrm{NA}}=1580$ for LiC$_6$. The calculated values are summarized in Table \ref{tab:raman}.

The last spectrum, d) in Fig \ref{fig:raman} shows an example with moderate intercalation, where a weak Fano is observed together with a symmetric Lorentzian shape. Here, the Fano shape is downshifted to $\omega_{\text{Fano}}=1514$ cm$^{-1}$ and its width is reduced to $\Gamma=25.4$ cm$^{-1}$. The value of the $q$ asymmetry parameter is $-1.29$. The Lorentzian peak is present at $\omega=1599$ cm$^{-1}$ with a width of $\Gamma=20.2$ cm$^{-1}$. Previous literature data suggest that the Lorentzian appearing at around $1600$ cm$^{-1}$ is a Stage-II compound phase \cite{Eklund1977,Nemanich1977,Eklund1980}, while the presence of the Fano suggests a de-intercalated LiC$_6$ phase \cite{ChaconPSSB2012}. Thus we identify this sample as a mixture of a de-intercalated Stage I and a Stage II phase.

The characteristic Raman modes of the alkali amides and lithium imide are not observed, their amount is thus negligible compared to the GICs.

\subsection{ESR results}

\begin{figure}[h!]
    \includegraphics*[width=\linewidth]{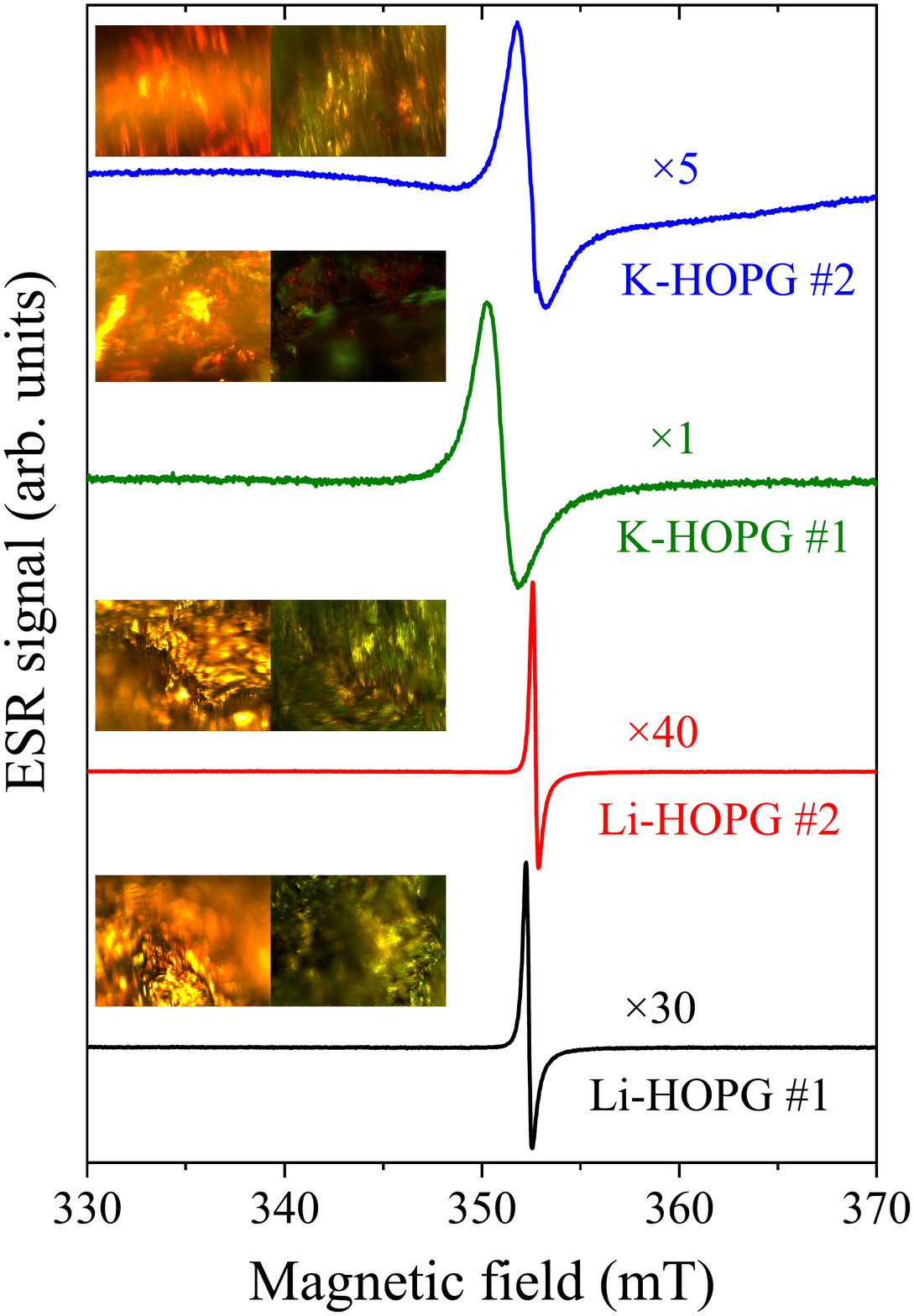}
    \caption{ESR spectra of the alkali intercalated HOPG samples prepared with the ammonia solution method. The materials exhibit a single Dysonian line near the $g$-factor of the free electron with a width compatible with previous literature observations \cite{Dresselhaus1981,Delhaes1977,Lauginie1980}. Inset shows bright and dark field photographs of the samples, where the dominant yellow (for Li) and red (for K) color is observable giving a visual proof of intercalation.}
    \label{fig:esr}
\end{figure}

ESR spectra of the prepared HOPG species are presented in Fig. \ref{fig:esr}. with insets showing bright field (BF) and dark field (DF) images.

The ESR spectra of the samples exhibit a characteristic, non-symmetric lineshape, which is identified as a Dysonian lineshape \cite{Dyson1955}. The presence of this lineshape is usually a clear indication of conducting electrons present in the system. It was shown that under certain circumstances (e.g. when the spin diffusion time is smaller than the spin relaxation time, this is the so-called "NMR-limit") Dysonian lines can be approximated with a weighted sum of an absorption and a dispersion Lorentzian \cite{Walmsley1996}. Here it is worth to note that charge neutral bulk graphite also presents a similar line with a linewidth of $\sim 0.29$ mT and $\sim 0.51$ mT for $B \perp c$ and $B \parallel c$, respectively \cite{Sercheli2002,Huber2004,MarkusGr2019}. Since the width is different for the intercalated species: $\sim 0.26$ mT for Li and $\sim 1.29$ mT for K, in a good agreement with previous literature values \cite{Dresselhaus1981,Delhaes1977,Lauginie1980}, thus, a successful bulk intercalation (at least down to the skin-depth, about $10$ microns \cite{FabianPRB2012}) is identified. Here, no angular-dependent measurements can be made, because the boiling ammonia makes the surface curly and ripply. Nevertheless, since the spin relaxation and $g$-factor anisotropy of the host graphite crystal is gradually reduced upon K intercalation \cite{MarkusPSSb2016}, the measured value can be treated as a mean value of the two different orientations. The $g$-factor of the prepared samples is close to that of free electron ($g_e=2.0023$), as expected from a metallic material. The spin relaxation mechanism of M$_{\text{am}}$-GICs is well understood in the framework of the Elliott--Yafet theory of spin-relaxation in metals \cite{Elliott1954,Yafet1963}, i.e. for Li the spin-orbit coupling is much weaker than for K, which results in a broader line for the latter. The parameters of the fitted curves are summarized in Table \ref{tab:samples}.

\begin{table}[h!]
    \begin{center}
        \begin{tabular}{cccc}
            \hline
            \hline
            Sample      & $g$-factor & $\Delta B$ (mT) & Phase  \\ \hline
            Li $\#1$    & $2.0029$   & $0.28$          & $27.8$ \\
            Li $\#2$    & $2.0024$   & $0.25$          & $32.5$ \\
            K $\#1$     & $2.0026$   & $1.33$          & $26.3$ \\
            K $\#2$     & $2.0027$   & $1.23$          & $20.7$ \\
            \hline
            \hline
        \end{tabular}
    \end{center}
    \caption{Fitted ESR parameters of the prepared samples. The $g$-factor of the prepared samples is close to the free electron's, the extracted linewidth is compatible with previous literature observations \cite{Dresselhaus1981,Delhaes1977,Lauginie1980,FabianPRB2012}. The Dysonian lineshape is approximated with a sum of two Lorentzians, where the weight is denoted with the phase.}
    \label{tab:samples}
\end{table}

The ESR data is in a good agreement with previous results, where the samples were prepared with the two zone vapor phase or with the immersion method \cite{Dresselhaus1981,Delhaes1977,Lauginie1980}. This means that our method provides samples of comparable quality to the standard ones, in a faster, commercially advantageous way. Moreover, there was no sign of emerging signal stemming from amides or imides.

\section{Summary}

We managed to improve the intercalation of carbonaceous materials in liquid ammonia in terms of reduced preparation time and quality comparable to conventional preparatory methods (two zone vapor phase and molten immersion). The reaction is demonstrated on graphite discs and the samples were investigated with optical microscopy, Raman, and ESR spectroscopy. The results suggest that the final materials are close to Stage-I GICs. We showed that it is possible to avoid the unwanted side reactions of the gaseous ammonia, which are suppressed and the amide-imide formation is blocked. Here, we wish to emphasize that our method involves the use of a vacuum system, which allows the preparation of the samples under totally inert, closed conditions. This not just results oxidation free materials, but also allows the reduce the amount of ammonia used. The prepared samples are sealed under vacuum or helium and thus readily measurable after the synthesis.

\section{Acknowledgement}
Work supported by the Hungarian National Research, Development and Innovation Office (NKFIH) Grant Nr. K119442, SNN118012 and 2017-1.2.1-NKP-2017-00001. The research reported in this paper was supported by the BME-Nanonotechnology FIKP grant of EMMI (BME FIKP-NAT). P.~Sz, B.~N., and L.~F. were supported by the Swiss National Science Foundation (Grant No. 200021 144419)

\bibliography{nh3intercal}



\end{document}